
%

\def\leaderdot{\leaders\hbox to 1 em {\hss.\hss}\hfill}

\dimen0= \parindent
\dimen1= \hsize \advance\dimen1 by -\dimen0

\dimen2=\baselineskip
\def\skiplines#1 { \dimen3=\dimen2 \multiply\dimen3 by #1 \vskip \dimen3}
\def\fullline{\hbox to \fullhsize}

\def\numpage{\baselineskip=24pt\fullline{\the\footline}}

\def\mathcedilla{\vtop{\hbox{c}{\kern0pt\nointerlineskip}
	         {\hbox{$\mkern-2mu \mathchar"0018\mkern-2mu$}}}}

\mathchardef\gq="0060
\mathchardef\dq="0027
\mathchardef\ssmath="19
\mathchardef\aemath="1A
\mathchardef\oemath="1B
\mathchardef\omath="1C
\mathchardef\AEmath="1D
\mathchardef\OEmath="1E
\mathchardef\Omath="1F
\mathchardef\imath="10
\mathchardef\fmath="0166
\mathchardef\gmath="0167
\mathchardef\vmath="0176


\def\colleft{\strut\kern.3em}
\def\colright{\kern0pt}

\def\figureh{\hbox to}

\def\m@th{\mathsurround=0pt}
\newif\ifdtpt
\def\displ@y{\openup1\jot\m@th
    \everycr{\noalign{\ifdtpt\dt@pfalse
    \vskip-\lineskiplimit \vskip\normallineskiplimit
    \else \penalty\interdisplaylinepenalty \fi}}}
\def\eqalignl#1{\,\vcenter{\openup1\jot\m@th
                \ialign{\strut$\displaystyle{##}$\hfil&
                              $\displaystyle{{}##}$\hfil&
                              $\displaystyle{{}##}$\hfil&
                              $\displaystyle{{}##}$\hfil&
                              $\displaystyle{{}##}$\hfil\crcr#1\crcr}}\,}
\def\eqalignnol#1{\displ@y\tabskip\centering \halign to \displaywidth{
                  $\displaystyle{##}$\hfil\tabskip=0pt &
                  $\displaystyle{{}##}$\hfil\tabskip=0pt &
                  $\displaystyle{{}##}$\hfil\tabskip=0pt &
                  $\displaystyle{{}##}$\hfil\tabskip=0pt &
                  $\displaystyle{{}##}$\hfil\tabskip\centering &
                  \llap{$##$}\tabskip=0pt \crcr#1\crcr}}
\def\leqalignnol#1{\displ@y\tabskip\centering \halign to \displaywidth{
                   $\displaystyle{##}$\hfil\tabskip=0pt &
                   $\displaystyle{{}##}$\hfil\tabskip=0pt &
                   $\displaystyle{{}##}$\hfil\tabskip=0pt &
                   $\displaystyle{{}##}$\hfil\tabskip=0pt &
                   $\displaystyle{{}##}$\hfil\tabskip\centering &
                   \kern-\displaywidth\rlap{$##$}\tabskip=\displaywidth
                   \crcr#1\crcr}}
\def\eqalignc#1{\,\vcenter{\openup1\jot\m@th
                \ialign{\strut\hfil$\displaystyle{##}$\hfil&
                              \hfil$\displaystyle{{}##}$\hfil&
                              \hfil$\displaystyle{{}##}$\hfil&
                              \hfil$\displaystyle{{}##}$\hfil&
                              \hfil$\displaystyle{{}##}$\hfil\crcr#1\crcr}}\,}
\def\eqalignnoc#1{\displ@y\tabskip\centering \halign to \displaywidth{
                  \hfil$\displaystyle{##}$\hfil\tabskip=0pt &
                  \hfil$\displaystyle{{}##}$\hfil\tabskip=0pt &
                  \hfil$\displaystyle{{}##}$\hfil\tabskip=0pt &
                  \hfil$\displaystyle{{}##}$\hfil\tabskip=0pt &
                  \hfil$\displaystyle{{}##}$\hfil\tabskip\centering &
                  \llap{$##$}\tabskip=0pt \crcr#1\crcr}}
\def\leqalignnoc#1{\displ@y\tabskip\centering \halign to \displaywidth{
                  \hfil$\displaystyle{##}$\hfil\tabskip=0pt &
                  \hfil$\displaystyle{{}##}$\hfil\tabskip=0pt &
                  \hfil$\displaystyle{{}##}$\hfil\tabskip=0pt &
                  \hfil$\displaystyle{{}##}$\hfil\tabskip=0pt &
                  \hfil$\displaystyle{{}##}$\hfil\tabskip\centering &
                  \kern-\displaywidth\rlap{$##$}\tabskip=\displaywidth
                  \crcr#1\crcr}}



\def\charlvmidlw#1#2{\,\vtop{\ialign{##\crcr
      #1\crcr\noalign{\kern1pt\nointerlineskip}
      $\hfil#2\hfil$\crcr}}\,}
\def\charlvlowlw#1#2{\,\vtop{\ialign{##\crcr
      $\hfil#1\hfil$\crcr\noalign{\kern1pt\nointerlineskip}
      #2\crcr}}\,}
\def\charlvmidup#1#2{\,\vbox{\ialign{##\crcr
      $\hfil#1\hfil$\crcr\noalign{\kern1pt\nointerlineskip}
      #2\crcr}}\,}
\def\charlvupup#1#2{\,\vbox{\ialign{##\crcr
      #1\crcr\noalign{\kern1pt\nointerlineskip}
      $\hfil#2\hfil$\crcr}}\,}

\def\vspce{\kern4pt} \def\hspce{\kern4pt}    

\def\emptybox{\vbox{\kern.7ex\hbox{\kern.5em}\kern.7ex}}
 \font\sevmi  = cmmi7              
    \skewchar\sevmi ='177          
 \font\fivmi  = cmmi5              
    \skewchar\fivmi ='177          
\font\tenmib=cmmib10
\newfam\bfmitfam

\textfont\bfmitfam=\tenmib
\scriptfont\bfmitfam=\sevmi
\scriptscriptfont\bfmitfam=\fivmi


\def\twodot{.\kern-0.1em.}

\def\paral{\mathrel{/\kern-.25em/}}
\def\grlo{\mathrel{\hbox{\lower.2ex\hbox{\rlap{$>$}\raise1ex\hbox{$<$}}}}}
\def\logr{\mathrel{\hbox{\lower.2ex\hbox{\rlap{$<$}\raise1ex\hbox{$>$}}}}}
\def\greq{\mathrel{\hbox{\lower1ex\hbox{\rlap{$=$}\raise1.2ex\hbox{$>$}}}}}
\def\loeq{\mathrel{\hbox{\lower1ex\hbox{\rlap{$=$}\raise1.2ex\hbox{$<$}}}}}
\def\grsim{\mathrel{\hbox{\lower1ex\hbox{\rlap{$\sim$}\raise1ex\hbox{$>$}}}}}
\def\losim{\mathrel{\hbox{\lower1ex\hbox{\rlap{$\sim$}\raise1ex\hbox{$<$}}}}}
\font\ninerm=cmr9
\def\uniset{\rlap{\ninerm 1}\kern.15em 1}

\def\emptysq{\mathbin{\vbox{\hrule\hbox{\vrule height1ex \kern.5em
                            \vrule height1ex}\hrule}}}
\def\emptyrect{\mathbin{\vbox{\hrule\hbox{\vrule height1ex \kern1em
                              \vrule height1ex}\hrule}}}
\def\rightonleftarrow{\mathrel{\hbox{\raise.5ex\hbox{$\rightarrow$}\ignorespaces
                                   \lower.5ex\hbox{\llap{$\leftarrow$}}}}}
\def\leftonrightarrow{\mathrel{\hbox{\raise.5ex\hbox{$\leftarrow$}\ignorespaces
                                   \lower.5ex\hbox{\llap{$\rightarrow$}}}}}

\def\bkB{{\rm I\kern-.17em B}}
\def\bkC{{\rm \kern.24em
            \vrule width.05em height1.4ex depth-.05ex
            \kern-.26em C}}
\def\bkD{{\rm I\kern-.17em D}}
\def\bkE{{\rm I\kern-.17em E}}
\def\bkF{{\rm I\kern-.17em F}}
\def\bkG{{\rm \kern.24em
            \vrule width.05em height1.4ex depth-.05ex
            \kern-.26em G}}
\def\bkH{{\rm I\kern-.22em H}}
\def\bkI{{\rm I\kern-.22em I}}
\def\bkJ{{\rm \kern.19em
            \vrule width.02em height1.5ex depth0ex
            \kern-.20em J}}
\def\bkK{{\rm I\kern-.22em K}}
\def\bkL{{\rm I\kern-.17em L}}
\def\bkM{{\rm I\kern-.22em M}}
\def\bkN{{\rm I\kern-.20em N}}
\def\bkO{{\rm \kern.24em
            \vrule width.05em height1.4ex depth-.05ex
            \kern-.26em O}}
\def\bkP{{\rm I\kern-.17em P}}
\def\bkQ{{\rm \kern.24em
            \vrule width.05em height1.4ex depth-.05ex
            \kern-.26em Q}}
\def\bkR{{\rm I\kern-.17em R}}
\def\bkT{{\rm \kern.24em
            \vrule width.02em height1.5ex depth 0ex
            \kern-.27em T}}
\def\bkU{{\rm \kern.30em
            \vrule width.02em height1.47ex depth-.05ex
            \kern-.32em U}}
\def\bkZ{{\rm Z\kern-.32em Z}}



%
\def\midfig#1x#2:#3#4{\midinsert
$$\vbox to #2{\hbox to #1{\special{ps:#3}\hfill}\vfill}$$\par
#4\par\endinsert}
\def\topfig#1x#2:#3#4{\topinsert
$$\vbox to #2{\hbox to #1{\special{ps:#3}\hfill}\vfill}$$\par
#4\par\endinsert}
\def\infig#1x#2:#3{
$$\vbox to #2{\hbox to #1{\special{ps:#3}\hfill}\vfill}$$}
\def\textfig#1x#2:#3{
$\vbox to #2{\hbox to #1{\special{ps:#3}\hfill}\vfill}$}
\input definit.tex
\magnification=1200
\baselineskip=22truept
\centerline{{\bf Eigenmodes of Decay and Discrete Fragmentation Processes}}
\bigskip
\centerline{B.G.Giraud and R.Peschanski}
\centerline{\it{Service Physique Th\'eorique, DSM-CE Saclay, F91191
Gif/Yvette, France}}

\centerline{and}

\centerline{Wei-hsing Ma}
\centerline{\it{Institute of High Energy Physics, Academia Sinica, 100039
Beijing, China}}

\bigskip
{\it Abstract}\ :\ Linear rate equations are used
to describe the cascading decay of
an initial heavy cluster into fragments. This representation
is based upon a triangular matrix of transition rates.
We expand the state vector of
mass multiplicities, which describes the process, into the biorthonormal
basis of eigenmodes provided by the triangular matrix. When the
transition rates have a scaling property in terms of mass ratios
at binary fragmentation vertices, we obtain solvable models with explicit
mathematical properties for the eigenmodes. A suitable continuous limit
provides an interpolation between
the solvable models. It gives a general relationship between the decay
products and the elementary transition rates.

\bigskip
In a previous paper$^{1)}$ we considered binary fragmentation
processes where any
fragment with mass number $k$ breaks into fragments with mass numbers
$j$ and $k-j,$ $j =1,2...k-1,$ with a probability $w_{j k}$
per unit of time.
It was assumed that $w_{j k}$ was time independent. By definition, $w_{jk}=0$
if $j \ge k$ and $w_{j k}$
is symmetric if $j$ is replaced by $k-j,$ naturally. (For technical reasons,
the coefficient $w_{j,2j}$ is twice the actual transition rate.)
Let $N_j(t)$ be the multiplicity of fragment $j$ at time $t$
in a process initiated from the decay of a cluster $A,$ namely
$N_j(0)=\delta_{jA}.$
The model under study is described by the following set of linear,
first order differential equations,
$$
{dN_j \over dt}= - c_j N_j + \sum_{k=j+1}^A w_{jk} N_k,\ j=1,...A,\ \
c_j=\sum_{\ell=1}^{j-1} {w_{\ell j} \over 2}\ ,\eqno (1)
$$
With components $N_j,\ j=1,...A,$ for a column vector
$|{\cal N}>,$ the system, Eqs.(1), boils down to
$d|{\cal N}> / dt = {\cal W |N>}$ with a triangular matrix $\cal W.$
The solution of Eqs.(1) is obviously $|{\cal N}>=\exp(-t{\cal W})|{\cal N}_0>,$
a sum of exponentials whose rates
of decay in time are the trivial eigenvalues of the triangular $\cal W,$
namely the diagonal matrix elements $-c_{\lambda}.$ The purpose of the present
letter is to take advantage of the expansion of
the evolution on the biorthonormal basis
of eigenmodes of ${\cal W},$
$$
|{\cal N}>=\sum_{\lambda=1}^A\ |\lambda> \exp(-t c_{\lambda}) <\tilde \lambda|
{\cal N}_0>= \sum_{\lambda=1}^A\ |\lambda> \exp(-c_{\lambda}t)\ Y^{\lambda}_A,
\eqno(2)
$$
where $Y^{\lambda}_A$ is the last component of the bra eigenvector
$<\tilde \lambda|.$ We are specially interested
in the analytical structure of the bra eigenvectors $<\tilde \lambda|,$
because of the synthetic information carried by
scalar products  $M_{\lambda}=<\tilde \lambda|{\cal N}>$.

\medskip
In our previous investigations$^{1)}$ we found numerical evidence
for specific exponents $q(\lambda)$ which govern a power-like
increase of the components of the bra eigenvectors,
$Y^{\lambda}_j \propto j^{q(\lambda)}.$
We took advantage of the fact that $\cal W$ has
a fixed bra eigenvector $<\tilde 1 \ |$, whose components
$Y^1_j=j,\ j=1,...A,$ express mass conservation. Indeed
the total mass is $M_1=<\tilde 1 \ |{\cal N}>=\sum_{j=1}^A j N_j$, with
$dM_1 / dt=<\tilde 1 \ |{\cal W}|{\cal N}>=0,$ since the first eigenvalue
$-c_1$ identically vanishes.
It was also convenient to define the ``mass
weighted multiplicity'' (MWM) vector $U$
with components $U_i=i N_i,$ whose evolution is governed by a matrix
${\cal V}$, with matrix elements ${\cal V}_{jk}=j{\cal W}_{jk}/k,$ hence
$dU/dt={\cal V}U.$
A numerical analysis of $U$ revealed scaling
properties of the fragmentation process. The present paper
attempts to give analytical proofs where only
numerical evidence was previously found.
In particular we are interested in those mathematical properties of the
bra $<\tilde \lambda|$
eigenvectors which may enlighten the
comprehension of the expansion, Eq.(2).

\medskip
Our main tools in this analysis are the following equations for the
components $Y^{\lambda}_i$ 
of the bra 
eigenvectors of ${\cal W,}$
$$
\sum_{i=\lambda}^{j-2}\ Y^{\lambda}_i\ w_{ij}+Y^{\lambda}_{j-1}\ w_{j-1,j}=
(c_j-c_{\lambda})\ Y^{\lambda}_j,\ \ j \ge \lambda \ . \eqno(3)
$$
Similar equations govern the components $X^{\lambda}_i$ of the
ket eigenvectors, naturally.
We keep in mind that 
$Y^{\lambda}_j=0$ if $j<\lambda$
and $X^{\lambda}_i=0$ if $i>\lambda.$
A suitable (biortho)normalization
results from the following boundary conditions,
$$
Y^{\lambda}_{\lambda}=X^{\lambda}_{\lambda}=1. \eqno(4)
$$ Note that the components of
the biorthonormal eigenvectors of the matrix ${\cal V}$ which
governs the MWM are easily deduced, as
${\cal Y}^{\lambda}_j=\lambda Y^{\lambda}_j/j$ and
${\cal X^{\lambda}_i}=iX^{\lambda}_i/\lambda.$

\medskip
It happens frequently that $c_j$ is a monotonically increasing function of $j.$
The physical interpretation is that the heavier the system,
the larger its total decay rate. Under such
a condition, the coefficients $c_j-c_{\lambda}$
in the right-hand side of Eqs.(3) are always positive when $j>\lambda.$
The left-hand sides of the same, Eqs.(3), depends only on positive coefficients
$w_{ij}$ and components $Y^{\lambda}_i$ of rank $i$ lower than $j,$\
$\lambda \le i \le j-1.$ It is easy to calculate $Y^{\lambda}_{\lambda+1}$
from $Y^{\lambda}_{\lambda}$ alone, which was chosen to be positive from the
normalization $Y^{\lambda}_{\lambda}=1.$ Thus
$Y^{\lambda}_{\lambda+1}$ is positive. A trivial recursive argument
then shows that
{\it all non vanishing components $Y^{\lambda}_j$
are positive.}\footnote{$^1$}{As a consequence of the positivity
of the bras and of biorthogonality,
the ket eigenvectors must show highly oscillatory components.
This demands a further, specific study.} 

\medskip
In the same way as in our earlier$^{1)}$ studies we first
select values of $w_{ij}$ which express scaling properties at vertices
of binary fragmentation. For this, the family of models we consider$^{2)}$
uses the form $w_{ij}=[f(i/j)+f(1-i/j)]\ j^{-a},$ and is
parametrized by a ``splitting'' function $f$ and an overall
exponent $a.$ For most of
the present analytical study, we restrict $a$ to the value
$a=1,$  because it allows a convenient continuous limit, valid for finite
values of $\lambda$ and large sizes $A$ of the matrix.
The value $a=1$ is for instance
the proper exponent for QCD fragmentation$^{3,4)}$
when we consider energy-momentum
fragmentation rather than mass fragmentation.

\medskip
A simplified, albeit generic form of $f$ is
parametrized by a second exponent, $b,$ in the form
$$
w_{ij}=\left[ \left({i \over j}\right)^{-b} +
\left(1-{i \over j}\right)^{-b} \right] j^{-a}. \eqno(5)
$$
The variable $x=i/j$\ (with its mandatory symmetry complement $1-x$)
is obviously the scaling mass ratio at the fragmentation vertex.
We note that the cases $a=1,\ b=0$ and $a=1,\ b=-1$ generate
the same eigenvectors,
since the corresponding matrices are strictly proportional, namely
$w_{ij}|_{b=0}=2\ w_{ij}|_{b=-1}=2/j.$

\medskip
For $a=1, b=-1,$ the simple nature of $w_{ij},$ namely $w_{ij}=1/j,$
reduces Eqs.(3) into
$$
{ j-1 \over j}(c_{j-1}-c_{\lambda})\ Y^{\lambda}_{j-1}+
{Y^{\lambda}_{j-1}\over j}=(c_j-c_{\lambda})\ Y^{\lambda}_j\ . \eqno(3a)
$$
Hence a recursion relation for the bra eigenvector components is found,
$$
{Y^{\lambda}_i \over Y^{\lambda}_{i-1}}={ i+\lambda-1 \over i-\lambda }\ ,
\eqno(6)
$$
and accordingly the bra eigenvector components are given by the formula,
$$
Y^{\lambda}_i = { (i+\lambda-1)! \over (i-\lambda)!\ (2\lambda-1) ! }\ .
\eqno(7)
$$
For large values of $j,$ the recursion relation expands in powers of $1/j,$
when $j$ is large, according to
$$
{Y^{\lambda}_{i+1}\over Y^{\lambda}_i}=1+{2\lambda-1\over j}+{\cal O}(1/j^2)\ ,
\eqno(8)
$$
which means that $Y^{\lambda}_i \propto i^{2\lambda-1}.$
This proves analytically
that the scalar products
$<\tilde \lambda|{\cal N}>=\sum_{i=\lambda}^AY^{\lambda}_iN_i$
of the bra eigenvectors with ${\cal N}$ express moments,
of order $q(\lambda)=2\lambda-1,$
of the multiplicity distributions, restricted to fragments
at least as heavy as $\lambda$. This justifies analytically,
for $a=1$ and $b=-1,0,$ the numerical evidence obtained
in our previous paper$^{1)}$.


\medskip
For $a=1,b=1$ the rates become $w_{ij}=j/[i(j-i)].$ The matrix ${\cal V}$
is simpler than ${\cal W},$
since then ${\cal V}_{ij}=1/(j-i)$ in the upper-right
triangle, which does not vanish. The diagonal is unchanged,
and so is, naturally, the lower-left, vanishing triangle. A straightforward
but slightly tedious argument shows that
$$
{\cal Y}^{\lambda}_j=
\lambda Y^{\lambda}_j/j={(j-1)! \over (j-\lambda)!(\lambda-1)!}\ .
\eqno(9) 
$$
One deduces from Eq.(9) that
$$
{ Y^{\lambda}_{j+1} \over Y^{\lambda}_j } = { (j+1) \over (j+1-\lambda) }=
1+{\lambda \over j}+{\cal O}(1/j^2), \eqno(10)
$$
hence now $q(\lambda)=\lambda.$ This makes a second case where the power law,
$Y^{\lambda}_j \propto j^{q(\lambda)},$ can be proved analytically.

\medskip
If $a$ is frozen to the value $a=1,$ it is tempting to interpolate between
the behavior $q(\lambda)=2\lambda-1$ found for $b=-1$ and that one,
$q(\lambda)=\lambda,$ valid for $b=1.$ Because of the identity of eigenvectors
for the $b=-1$ and $b=0$ cases, a parabolic interpolation
$$
q_{\rm int}(\lambda) = { \lambda (17-4b'^2) + 4b'^2-9 \over 8}, \ \ \
b' \equiv b+{1\over 2},
\eqno(11)
$$
is found.

\medskip
This interpolation turns out to be consistent with
the continuous limit condition
considered earlier, see Eq.(9) in reference$^{1)},$
$$
\eqalign{
c_{\lambda} =& \ {\lambda^{b-1} \over 2} \sum_{i=1}^{\lambda-1}
\left[i^{-b}+(\lambda-i)^{-b}\right] =
\ \lambda^{b-1}\sum_{i=1}^{\lambda-1}i^{-b} \cr
=& {1\over 2}\int_0^1 dx \left[ 1-x^q-(1-x)^q \right]
\left[ x^{-b}+(1-x)^{-b} \right] =
\int_0^1 dx \left[ 1-x^q-(1-x)^q \right] x^{-b}, \cr
} \eqno(12) 
$$
where we take advantage of the symmetry $w_{ij}=w_{j-i,j}.$ We can
use this continuous
limit, Eq.(12), to estimate $q$ as a function of $\lambda$ when $c_{\lambda}$
is known, whether analytically or numerically. The identity of the cases $b=0$
and $b=-1$ for Eq.(12) is transparent.
Also, for any value of $b,$ the pair $\lambda=1, q=1$ solves obviously Eq.(12).
Then, for $b=0,$ one gets at once from Eq.(12)
the condition $(\lambda-1)/\lambda=1-2/(q+1),$ hence $q(\lambda)=2\lambda-1,$
in agreement with Eq.(8). Furthermore for $b=1,$ with the
ansatz $q(\lambda)=\lambda,$ the recursive relation, derived from Eq.(12),
$$
c_{\lambda+1}-c_{\lambda} = {1 \over \lambda} = \int_0^1 dx \
{ x^{\lambda}+(1-x)^{\lambda}-x^{\lambda+1}-(1-x)^{\lambda+1} \over x},
$$
is easy to verify.

\medskip
We show on Fig.1 the plots (full lines)
of the values of $q(2),$ $q(5)$ and $q(8),$
respectively, obtained numerically from Eq.(12) for $-1.5 \losim b \losim 1.5.$
The same Fig.1 shows (dotted lines)
the corresponding interpolation predictions
$q_{\rm int}(\lambda)$ from Eq.(11). As long as
$-1.2 \losim b \losim 1.2,$ the continuous limit results and the parabolic
interpolations do not differ by more than 4\%.
Significant deviations between such estimates occur, however,
when $ 1.2 \losim |b| .$ It may be pointed out that the
continuous limit estimates are expected to be more rigorous.
Finally on Fig.2 we show, for e.g. $b=-0.5,$ $b=0.5$ and $b=1.5,$
that, when provided by Eq.(12), the behavior of $q$ as a function
of $\lambda$ remains essentially linear. Namely a parametrization
$q(\lambda) \simeq \alpha(b) \lambda + \beta(b)$ is reasonable,
despite the fact that $\alpha(b)$ and $\beta(b)$ are more complicated
than the quadratic forms present in Eq.(11).

\medskip
In conclusion, there is a firm ground for the validity of the
power law which governs the bra eigenvector components
and the corresponding interpretation in terms of moments
of the multiplicity vector ${\cal N}.$ The apparent simplicity of the
empirical law $q(\lambda) \simeq \alpha(b) \lambda + \beta(b),$
see for instance Eq.(11), expresses a non trivial relationship between very
different quantities, namely experimentally observable moments on one hand
and, on the other hand, much less observable parameters $b$ of the
microscopic mechanism of binary fragmentation.

\medskip \noindent
{\it Acknowledgments\ :}\  One of the authors (W-X Ma) thanks the
French Atomic Energy Commission for a stay at Service de Physique Th\'eorique,
Saclay, where this work was performed.

\medskip
\centerline{{\bf References}}

\smallskip \noindent
1) B.G. Giraud and R. Peschanski, {\it Phys.Lett.}{\bf B315},452(1993)

\smallskip \noindent
2) Z. Cheng and S. Redner, {\it J.Phys.A:Math.Gen.}{\bf 23},1233(1990);
E.D. Mc Grady and Robert M. Ziff, {\it Phys.Rev.Lett.}{\bf 58},892(1987)

\smallskip \noindent
3) G. Altarelli and G. Parisi, {\it Nucl.Phys.}{\bf 126},297(1977);
V.N. Gribov and L.N. Lipatov,
{\it Sov.Journ.Nucl.Phys.}{\bf 15},438,675(1972);
For a review and references, see Y.L. Dokshitzer, V.A. Khoze, A.H. Mueller and
S.I. Troyan, {\it Basics of perturbative QCD} (J. Tran Than Van ed.
Editions Fronti\` eres, France, 1991)

\smallskip \noindent
4) P. Cvitanovic, P. Hoyer and K. Zalewski,
{\it Nucl.Phys.}{\bf B176},429(1980)

\smallskip \noindent
{\bf Figure Captions}

\smallskip \noindent
Fig.1 : Behavior of the ``eigenexponents''
$q({\lambda})$ for the 2nd, 5th and 8th bra eigenvectors, as functions of the
parameter $b$ (model with scaling properties for fragmentation vertices).
Full lines: continuous limit predictions. Dotted lines: quadratic
interpolations between the exact values, known analytically for $b=-1,0,1.$

\smallskip \noindent
Fig.2 : Strongly linear behavior of $q({\lambda})$ as a function of
${\lambda}$, for different values of $b.$ Full lines are drawn to guide the
eye between integer values of $\lambda.$

\bye